\documentstyle[sprocl]{article}
\def\NPB{{\em Nucl.Phys.}B}
\def\NPA{{\em Nucl.Phys.}A}

\def\PRL{{\em Phys. Rev. Lett.}}
\def\PRD{{\em Phys. Rev.} D}

\def\bi{\begin{itemize}}
\def\ei{\end{itemize}}
\def\be{\begin{equation}}
\def\ee{\end{equation}}
\def\bea{\begin{eqnarray}}
\def\eea{\end{eqnarray}}
\begin{document}
\vspace*{-2.cm}
\title{FERMIONS ON THE LIGHT-FRONT} 
\author{M.~BURKARDT}
\address{Department of Physics,New Mexico State University \\
Las Cruces, New Mexico 88003, U.S.A.}
\maketitle
\begin{abstract}
Issues that are specific for formulating fermions in light-cone 
quantization are discussed. Special emphasis is put on
the use of parity invariance in the non-perturbative renormalization
of light-cone Hamiltonians.
\end{abstract}
Light-front (LF) quantization is the most physical approach to
calculating parton distributions on the basis of QCD~\cite{adv}.
Before one can formulate QCD with quarks, it is necessary that
one understands how to describe fermions in this framework.
This in turn requires that one addresses the following issues
\bi
\item How is spontaneous symmetry breaking (chiral symmetry!)
manifested in the LF framework, where the vacuum
appears to be trivial?
\item Is it possible to preserve current conservation and
parity invariance in this framework?
\item How does one formulate fermions on the transverse lattice,
which seems to be a very promising approaches to pure glue
LFQCD \cite{bardeen,dalley}?\footnote{Because of lack of space,
this important issue could not be discussed in these notes.}
\ei
\section{Spontaneous Symmetry Breaking}
The first of the above issues has been addressed very often in
the past and we will restrict ourselves here to a brief
summary.\footnote{See for example Ref. \cite{adv} for a more
extensive discussion on this question.} In the LF framework, 
non-trivial vacuum structure can reside only in zero-modes
($k^+=0$ modes).
Since these are high-energy modes (actually infinite energy
in the continuum) one often does not include them as explicit
degrees of freedom but assumes they have been integrated out,
leaving behind an effective LF-Hamiltonian. The important points
here are the following. If the zero-mode sector involves
spontaneous symmetry breaking, this manifests itself as
explicit symmetry breaking for the effective Hamiltonian.
In general, these effective LF Hamiltonians
thus have a much richer operator structure than the canonical
Hamiltonian. Therefore, compared to a conventional Hamiltonian
framework, the question of the vacuum has been shifted from
the states to the operators and it should thus be clear that
the issues of renormalization and the vacuum are deeply
entangled in the LF framework.

\section{Current Conservation}
Despite widespread confusion on this subject, vector current 
conservation (VCC) is actually not a problem in the LF 
framework. Many researchers avoid the subject
of current conservation because the divergence of the vector 
current
\be
q_\mu j^\mu (q) = q^-j^+ + q^+j^- -
{\vec q}_\perp {\vec j}_\perp
\ee
involves $j^-$, which is quadratic in the constrained 
fermion spinor component
\be
j^-= \psi^\dagger_{(-)}\psi_{(-)},
\ee
and thus $j^-$ contains quartic interactions making it
at least as difficult to renormalize as the Hamiltonian.

We know already from renormalizing $P^-$ that the
canonical relation between $\psi_{(-)}$ and $\psi_{(+)}$
is in general not preserved in composite operators 
(such as $\bar{\psi}\psi$). It is therefore clear that
a canonical definition for $j^-$ will in general violate
current conservation since it does not take into account
integrating out zero-modes and other high-energy degrees
of freedom.

However, in LF gauge $A^+=0$, $j^-$ does not enter the
Hamiltonian and therefore one can address its definition
separately from the construction of the Hamiltonian.
In fact, it is very easy to find a pragmatic definition
which obviously guarantees manifest current conservation, 
namely
\bea
j^-(q^+) &=& -\frac{1}{q^+} \left[ P^-, j^+(q^+)\right]
\quad \quad \quad \quad \quad \quad (1+1) \label{eq:j-def}\\
j^-(q^+,{\vec q}_\perp) &=& -\frac{1}{q^+} \left\{
\left[ P^-, j^+(q^+,{\vec q}_\perp)\right]- 
{\vec q}_\perp{\vec j}_\perp
(q^+,{\vec q}_\perp)\right\}
\quad \quad \quad (3+1)\nonumber
\eea
in 1+1 and 3+1 dimensions respectively.
The corresponding expressions in coordinate space 
$(x^-,{\vec x}_\perp)$ can be obtained by Fourier transform.
In summary,
\bi
\item Most importantly, VCC is no
problem in LF quantization and is manifest at the operator 
level, provided $j^-$ is {\it defined} using Eq. 
(\ref{eq:j-def}).
\item Since VCC can easily be made
 manifest (by using the above construction!), there is no point 
in {\em testing} its validity and it cannot be used as a 
non-perturbative renormalization condition either.
\item For a non-interacting theory, Eq. (\ref{eq:j-def}) 
reduces to the canonical definition of $j^-$, but in general
this is not the case when $P^-$ contains interactions or even
non-canonical terms. \ei
Note that (as so often on the LF) $q^+$
appears in the denominator of Eq. (\ref{eq:j-def}).
Therefore, as usual, one should be very careful while
taking the $q^+\rightarrow 0$ limit and while drawing
any conclusions about this limit.
An example of this kind are the pair creation terms in $j^-$.
Naively they do not contribute for $q^+\rightarrow 0$,
since the $q\bar{q}$ pair emanating from $j^-$ necessarily
carries positive $q^+$. However, since $j^-$ often has very 
singular matrix elements for $q^+$, such seemingly vanishing
terms nevertheless survive the $q^+\rightarrow 0$ limit,
which often leads to confusion. For an early example of this
kind see Ref. \cite{mb:1+1}. A more recent discussion can be
found in Ref. \cite{recent}.
\section{Parity Invariance}
\subsection{General Remarks}
A parity transformation, $x^0\stackrel{P}{\longrightarrow} 
x^0$, ${\vec x}\stackrel{P}{\longrightarrow}-{\vec x}$
leaves the quantization hyperplane ($x^0=0$) 
in equal time (ET) quantization invariant
and therefore the parity operator is a kinematic
operator in such a framework. It is thus very easy
to ensure that parity is a manifest symmetry in ET
quantization by tracking parity at each step in a
calculation.
The situation is completely different in the LF framework,
where the same parity transformation exchanges LF-`time' 
($x^+\equiv x^0+x^3$) and space ($x^-\equiv x^0-x^3$) 
directions, i.e.
\be
x^+ \stackrel{P}{\longleftrightarrow}x^-
\ee
and therefore the quantization hyperplane 
$x^+=0$ is {\it not} invariant. Hence, the
parity operator is a dynamical operator on the LF and, 
except for a free field theory, it is probably impossible to
write down a simple expression for it in terms of quark
and gluon field operators. Thus, parity invariance
is not a manifest symmetry in this framework.
Note that the situation
is the other way round for the boost operator (kinematic and
manifest on the LF, dynamical and non-manifest in ET).
It thus depends on the physics application one is interested
in and the symmetries that one considers the most important
ones for that particular physics problem, which framework is
preferable.

For most applications of LF quantization, the lack of manifest
parity actually does not constitute a problem --- in fact, one
can view it as an opportunity rather than a problem.
The important point here is the following:
due to the lack of manifest covariance in a Hamiltonian 
formulation, LF Hamiltonians in general contain more parameters
than the corresponding Lagrangian. Parity invariance may be
very sensitive to some of these parameters. 

In order to illustrate this important point, let us consider
the example of a 1+1 dimensional Yukawa model \cite{parity}
\begin{equation}
{\cal L}=\bar{\psi}\left( i\not \!\!\partial 
-m_F-g\phi \gamma_5 \right)\psi -\frac{1}{2}
\phi\left( \Box +m_B^2\right)\phi .
\end{equation}
This model actually has a lot in common with the kind of interactions that
appear when one formulates QCD (with fermions) on a transverse lattice 
\cite{hala}.

The main difference between scalar and Dirac fields in the LF formulation is
that not all components of the Dirac field are dynamical: multiplying the
Dirac equation
\begin{equation}
\left( i\not \!\!\partial 
-m_F-g\phi \gamma_5\right)\psi =0
\end{equation}
by $\frac{1}{2}\gamma^+$ yields a constraint equation (i.e. an
``equation of motion'' without a time derivative)
\begin{equation}
i\partial_-\psi_{(-)}=\left(m_F+g\phi\gamma_5\right)\gamma^+\psi_{(+)}
,
\label{eq:constr}
\end{equation}
where
$
\psi_{\pm}\equiv \frac{1}{2}\gamma^\mp \gamma^\pm \psi .
$
For the quantization procedure, it is convenient to eliminate
$\psi_{(-)}$, using
\be
\psi_{(-)} = \frac{\gamma^+}{2i\partial_-}\left(m_F+g\phi\gamma_5\right) 
\psi_{(+)}  
\ee
from the classical Lagrangian before imposing
quantization conditions, yielding
\begin{eqnarray}
{\cal L}&=&\sqrt{2}\psi_{(+)}^\dagger i\partial_+ \psi_{(+)}
-\phi\left( \Box +m_B^2\right)\phi
-\psi^\dagger_{(+)}\frac{m_F^2}{\sqrt{2}i\partial_-}
\psi_{(+)}
\label{eq:lelim}
\\
&-&\psi^\dagger_{(+)}\left(
g\phi
\frac{m_F\gamma_5}{\sqrt{2}i\partial_-}
+\frac{m_F\gamma_5}{\sqrt{2}i\partial_-}g\phi\right)
\psi_{(+)}
-\psi^\dagger_{(+)}g\phi\frac{1}{\sqrt{2}i\partial_-}
g\phi\psi_{(+)} .
\nonumber
\end{eqnarray}
The rest of the quantization procedure very much resembles the procedure
for self-interacting scalar fields.

The above canonical Hamiltonian contains a kinetic term for the fermions,
a fermion boson vertex and a fermion 2-boson vertex. 
While the couplings of these three terms in the canonical Hamiltonian
depend only on two independent parameters ($m$ and $g$), it turns out
that these terms are renormalized independently from each other once
zero-mode and other high-energy degrees of freedom are integrated out.
More explicitly this
means that one should make an ansatz for the renormalized LF Hamiltonian
density of the form
\begin{eqnarray}
{\cal P}^-&=&
\frac{m_B^2}{2}\phi^2
+\psi^\dagger_{(+)}\frac{c_2}{\sqrt{2}i\partial_-}
\psi_{(+)}
+c_3\psi^\dagger_{(+)}\left(
\phi
\frac{\gamma_5}{\sqrt{2}i\partial_-}
+\frac{\gamma_5}{\sqrt{2}i\partial_-}g\phi\right)
\psi_{(+)} \nonumber\\
&+&c_4\psi^\dagger_{(+)}\phi\frac{1}{\sqrt{2}i\partial_-}
\phi\psi_{(+)} ,
\label{eq:pren}
\end{eqnarray}
where the $c_i$ do not necessarily satisfy the canonical relation
$c_3^2=c_2c_4$. However, this does not mean that the $c_i$ 
are completely
independent from each other. In fact, Eq.(\ref{eq:pren}) will 
describe the Yukawa model only for specific combinations of
$c_i$. It is only that
we do not know the relation between the $c_i$.
\footnote{Coupling coherence \cite{brazil} does not help to 
determine the finite part (integration constant) for the mass 
if the original Lagrangian contains a nonzero bare mass!}

Thus the bad new is that the number of parameters in the LF Hamiltonian
has increased by one (compared to the Lagrangian). The good news is that
a wrong combination of $c_i$ will in general give rise to a parity 
violating theory: formally this can be seen in the weak coupling
limit,
where the correct relation ($c_3^2=c_2c_4$) follows from a covariant
Lagrangian. Any deviation from this relation can be described on
the level of the Lagrangian (for free massive fields, equivalence
between LF and covariant formulation is not an issue)
by addition of a term of the form $\delta {\cal L} =
\bar{\psi}\frac{\gamma^+}{i\partial^+}\psi$, which is 
obviously 
parity violating, since parity transformations result in
$A^\pm \stackrel{P}{\rightarrow} A^\mp$ for Lorentz vectors 
$A^\mu$; i.e. $\delta {\cal L}
\stackrel{P}{\rightarrow} \bar{\psi}\frac{\gamma^-}{i\partial^-}\psi \neq \delta {\cal L}$.
This also affects physical observables, as can be seen by
considering boson fermion scattering
in the weak coupling limit of the Yukawa model.
At the tree level, there is an instantaneous contact 
interaction, which is proportional to $\frac{1}{q^+}$. 
The (unphysical) singularity at $q^+=0$ is canceled by
a term with fermion intermediate states, which contributes
(near the pole) with an amplitude 
$\propto -\frac{m_v^2}{m_{kin}^2}\frac{1}{q^+}$.
Obviously, the singularity cancels iff $m_V=m_{kin}$.
Since the singularity involves the LF component $q^+$,
this singular piece obviously changes under parity.
This result is consistent with the fact that there is no 
zero-mode induced renormalization of $m_{kin}$ and thus 
$m_V=m_{kin}$ at the tree level.
This simple example clearly demonstrates that a `false'
combination of $m_V$ and $m_{kin}$ leads to violations of
parity for a physical observable, which is why imposing
parity invariance as a renormalization condition may help
reduce the dimensionality of coupling constant space.

\subsection{Parity Sensitive Observables that ``don't work''}
Of course there are an infinite number of parity sensitive 
observables, but not all of them are easy to evaluate 
non-perturbatively in the Hamiltonian LF framework. 
Furthermore, 
we will see below that some relations among observables, 
which seem to be sensitive to parity violations, are 
actually `protected'
by manifest symmetries, such as charge conjugation, or by VCC.

From the brief discussion of the (to-be-canceled) singularity
above it seems that the most sensitive observable to look for
violations of parity in QCD would be Compton scattering cross 
sections between quarks and gluons (or the corresponding
fermions and bosons in other field theories) because there one 
could tune the external momenta such that the potential 
singularity enters with maximum strength. However, this is
not a very good choice: first of
all non-perturbative scattering amplitudes are somewhat
complicated to construct on the LF. Secondly, quarks and gluons
are confined particles, which makes $qg$ Compton scattering
an unphysical process. 

A much better choice are any matrix elements between
bound states. Bound states are non-perturbative and 
all possible momentum transfers occur in their
time evolution. Therefore, any parity violating sub-amplitude would 
contribute at some point and would therefore affect 
physical observables. Secondly, since one of the primary goal
of LFQCD is to explore the non-perturbative spectra and structure
of hadrons, matrix elements in bound states are the kind of 
observables for which the whole framework has been tailored.

One conceivable set of matrix elements are those of the
vector current operator $j^\pm$. For example, consider
the vacuum to meson matrix elements
\bea
\langle 0 | j^+ |n,p\rangle &=& p^+ f_n^{(+)} \nonumber\\
\langle 0 | j^- |n,p\rangle &=& p^- f_n^{(-)}.
\label{eq:fpm}
\eea
By boost invariance, the couplings defined in Eq. (\ref{eq:fpm})
must be independent of the momenta.
Obviously, parity invariance requires $|f_n^{(+)}| =
|f_n^{(-)}|$. However, this relation also follows from
current conservation $0=p^-\langle 0 | j^+ |n,p\rangle 
+p^+\langle 0 | j^- |n,p\rangle $, which makes it a useless
relation for the purpose of parity tests.

Similar statements can be made about elastic formfactors
but we will omit the details here. The basic upshot is that
the same relations between the matrix elements of $j^\pm$
that arise from parity invariance can often
also be derived using
only VCC, i.e. such relations are
in general `protected' by VCC

One may also consider non-conserved currents, such as the
axial vector current. However, there one would have to face
the issue of defining the `minus' component before one can
test any parity relations. Parity constraints are probably 
very helpful in this case when constructing the minus 
components, but then those relations can no longer be used to 
help constrain the coefficients in the LF-Hamiltonian.

Another class of potentially useful operators consists of the
scalar and pseudoscalar densities $\bar{\psi}\psi$ and
$\bar{\psi} \gamma_5 \psi$. Obviously, if parity is conserved then
at most one of the two couplings
\bea
f_S&=& \langle 0 | \bar{\psi}\psi | n\rangle \nonumber\\
f_P&=& \langle 0 | \bar{\psi}\gamma_5\psi | n\rangle
\eea
can be nonzero at the same time, since a state $|n\rangle$
cannot be both scalar and pseudoscalar.
What restricts the usefulness of this criterion is the fact
that the same `selection rule' also follows from
charge conjugation invariance ($\bar{\psi}$ and
$\bar{\psi} \gamma_5 \psi$ have opposite charge parity!)
which is a manifest symmetry on the LF. Therefore, only for
theories with two or more flavors, where one can consider
operators such as $\bar{u}s$ and $\bar{u}\gamma_5 s$, does one
obtain parity constraints that are not protected by charge
conjugation invariance. But even then one may have to face
the issue of how to define these operators. Parity may of 
course be used in this process, but then it has again been
`used up' and one can no longer use these selection rules to
test the Hamiltonian.

In summary, many parity relations are probably very useful
to determine the LF representation of the operators involved,
but may not be very useful to determine the Hamiltonian.

\subsection{A useful observable to test parity}

We have seen above that relations between $j^+$ and $j^-$ are
often protected by vector current conservation and are therefore not
very useful as parity tests. A much more useful parity test can 
actually be obtained by considering the `plus'-component only:
Let us now consider a vector form factor,
\begin{equation}
\langle p^\prime, n| j^\mu | p, m \rangle
\stackrel{!}{=} \varepsilon^{\mu \nu}q_\nu F_{mn}(q^2),
\label{eq:form}
\end{equation}
where $q=p^\prime -p$, between states of opposite parity. 
When writing the r.h.s. in terms of
one invariant form factor, use was made of both vector current
conservation and parity invariance. A term proportional
to $p^\mu + {p^\prime}^\mu$ would also satisfy current conservation,
but has the wrong parity. A term proportional to 
$\varepsilon^{\mu \nu}\left(p_\nu + p_\nu^\prime\right)$ 
has the right parity,
but is not conserved and a term proportional to $q^\mu$ is both
not conserved and violates parity. Other vectors do not exist for
this example.
The Lorentz structure in Eq. (\ref{eq:form}) has nontrivial
implications even if we consider only the ``plus'' component, 
yielding
\begin{equation}
\frac{1}{q^+}\langle p^\prime, n| j^+ | p, m \rangle
=F_{mn}(q^2).
\label{eq:formplus}
\end{equation}

\begin{figure}
\unitlength1.cm
\begin{picture}(10,12)(0,1.)
\includegraphics{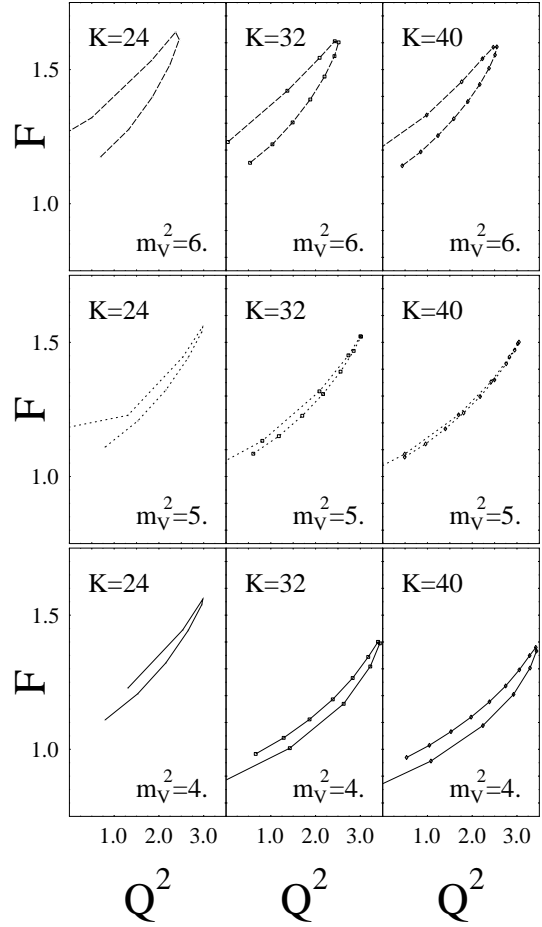}
\end{picture}
\caption{
Inelastic transition form factor (\protect\ref{eq:formplus}) between the
two lightest meson states of the Yukawa model, calculated
for various vertex masses $m_v$ and for various DLCQ parameters
$K$. The physical masses for the fermion and the scalar meson
have been renormalized to the values 
$\left(m_F^{phys}\right)^2=\left(m_F^{phys}\right)^2=4$.
All masses and momenta are in units of
$\protect\sqrt{\lambda} = \protect\sqrt{c_4/2\pi}$. In this example,
only for $m_V^2\approx 5$ one obtains a form factor that is a unique
function of $Q^2$, i.e. only for $m_V^2\approx 5$, the result is
consistent with Eq. (\protect\ref{eq:formplus}). Therefore, only for
this particular value of the vertex mass, is the matrix element of the current operator consistent with both parity and current conservation.
}
\label{fig:parity}
\end{figure}
That this equation implies nontrivial constraints can be seen
as follows: as a function of the longitudinal momentum transfer
fraction $x\equiv q^+/p^+$, the invariant momentum transfer 
reads ($M_m^2$ and $M_n^2$ are the invariant masses of the
in and outgoing meson)
\begin{equation}
q^2=x\left(M_m^2 -\frac{M_n^2}{1-x}\right).
\end{equation}
This quadratic equation has in general, for a given value of $q^2$,
two solutions for $x$ which physically correspond to hitting the
meson from the left and right respectively. The important point is that
it is not manifestly true that these two values of $x$ in Eq. 
(\ref{eq:formplus}) will give the same value for the form factor,
which makes this an excellent parity test. 

In Ref. \cite{parity}, the coupling as well as the 
physical masses of both the fermion and the lightest boson where
kept fixed, while the ``vertex mass'' was tuned (note that this
required re-adjusting the bare kinetic masses). Figure \ref{fig:parity}
shows a typical example, where the calculation of the
form factor was repeated for three values of the DLCQ parameter K
(24, 32 and 40) in order to make sure that numerical approximations
did not introduce parity violating artifacts.

For a given physical mass and boson-fermion coupling, there
exists a ``magic value'' of the vertex mass and only for 
this value one finds that the parity condition 
(\ref{eq:formplus}) is satisfied over the whole range of $q^2$
considered. This provides a strong self-consistency check, since
there is only one free parameter, but the parity condition is not
just one condition but a condition for every single value of
$q^2$ (i.e. an infinite number of conditions). In other words,
keeping the vertex mass independent from the kinetic mass is not
only necessary, but also seems sufficient in order to properly 
renormalize Yukawa$_{1+1}$.

In the above calculation, it was sufficient to work with
a vertex mass that was just a constant.
However, depending on the interactions and the cutoffs employed,
it ma be necessary to introduce counter-term functions\cite{brazil}.
But even in cases where counter-term functions
need to be introduced, parity constraints may be very helpful
in determining the coefficient-functions for those more
complex effective LF Hamiltonians non-perturbatively.


\section*{References}
\begin{thebibliography}{99}
\bibitem{adv} M.~Burkardt, {\em Advances Nucl. Phys.} 
{\bf 23}, 1 (1996).
\bibitem{bardeen} W.~A.Bardeen et al., 
\PRD {\bf 21}, 1037 (1980).
\bibitem{dalley} S. Dalley and B. van de Sande, \PRL {\bf 82},
1088 (1999); \PRD {\bf 59}, 065008 (1999).
\bibitem{mb:1+1} M.Burkardt, \NPA {\bf 504}, 762 (1989);
\NPB {\bf 373}, 613 (1992).
\bibitem{recent} S.J.~Brodsky and D.S.~Hwang, \NPB {\bf 543}, 239 (1998);
H.-M.~Choi and C.R.~Ji, \PRD {\bf 58}, 071901 (1998).
\bibitem{parity} M.~Burkardt, \PRD {\bf 54}, 2913 (1996).
\bibitem{hala} M. Burkardt and H. El-Khozondar, \PRD 
{\bf 60},
054504 (1999).
\bibitem{brazil} R.J. Perry, lectures given at NATO Advanced 
Study Institute on Confinement, Duality and Nonperturbative
Aspects of QCD,
Cambridge, U.K., 1997, hep-th/9710175.
\end{thebibliography}
\end{document}